\documentclass[sort&compress,final]{aipproc}\layoutstyle{6x9}
\usepackage{amsmath} 

\begin{document}\title{Effective Continuum Thresholds for
Quark--Hadron Duality in Dispersive Sum Rules}

\author{Wolfgang LUCHA}{address={Institute for High Energy Physics,
Austrian Academy of Sciences,\\Nikolsdorfergasse 18, A-1050
Vienna, Austria}}\author{Dmitri MELIKHOV}{address={Institute for
High Energy Physics, Austrian Academy of
Sciences,\\Nikolsdorfergasse 18, A-1050 Vienna,
Austria},altaddress={Faculty of Physics, University of Vienna,
Boltzmanngasse 5, A-1090 Vienna, Austria}}\author{Silvano
SIMULA}{address={INFN, Sezione di Roma III, Via della Vasca Navale
84, I-00146 Roma, Italy}}

\begin{abstract}Modifying the standard approaches to
nonperturbative QCD based on Borel-transformed dispersive sum
rules by allowing the effective continuum thresholds required for
the implementation of quark--hadron duality to depend on the Borel
parameters and on any relevant momentum promises to entail higher
accuracy and reliable error estimates for the extracted
predictions of hadron features.\end{abstract}\maketitle

\section{Incentive: improvement of QCD sum rules \cite{LMS:PRD76,
LMS:QCD@Work07,LMS:YaF,LMS:PLB657,LMS:Hadron07,LMS:PLB671,DM:PLB671,
LMS:C8p,LMS:C8t,LMS:PRD79,LMS:JPG37,LMS:PRD80,LMS:PLB687,LMS:YaF2010}}
Within the method of QCD sum rules, \`a la
Shifman--Vainshtein--Zakharov, the concept of quark--hadron
duality is usually implemented by {\em assuming\/} that above
specific continuum thresholds the contributions to suitably
defined correlators of interpolating currents at the level of the
QCD degrees of freedom equal those at the level of the hadronic
bound states. Our approach \cite{LMS:PRD76,
LMS:QCD@Work07,LMS:YaF,LMS:PLB657,LMS:Hadron07,LMS:PLB671,DM:PLB671,
LMS:C8p,LMS:C8t,LMS:PRD79,LMS:JPG37,LMS:PRD80,LMS:PLB687,LMS:YaF2010}
seeks to quantify the uncertainty induced by such an {\em
approximation\/} and to improve the accuracy of all predictions by
allowing our threshold to depend on the involved momenta and
parameters introduced upon application of Borel transformations.
The Borel transformation to a new variable, the Borel parameter
(called $T$ or $\tau$ hereafter), serves to {\em remove\/}
subtraction terms and {\em suppress\/} excitation and continuum
contributions.

\begin{figure}[ht]\begin{tabular}{cc}
\includegraphics[height=.204\textheight,scale=1]{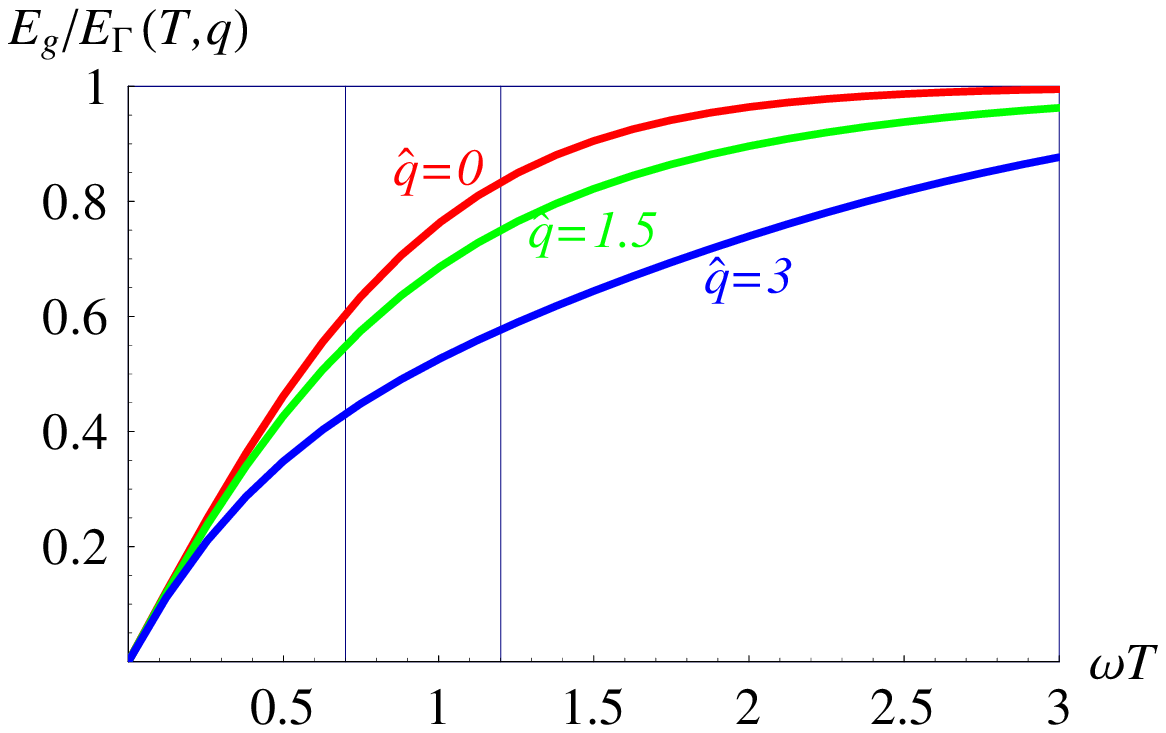}&
\includegraphics[height=.204\textheight,scale=1]{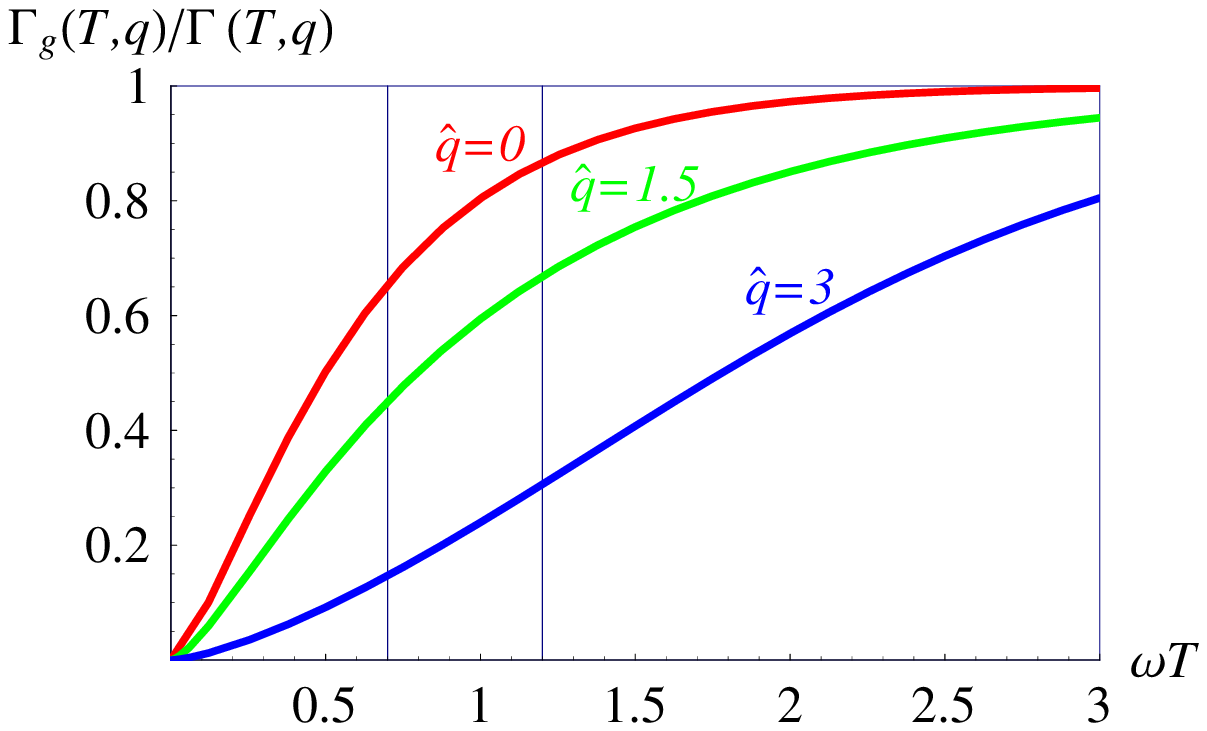}\\
\includegraphics[height=.204\textheight,scale=1]{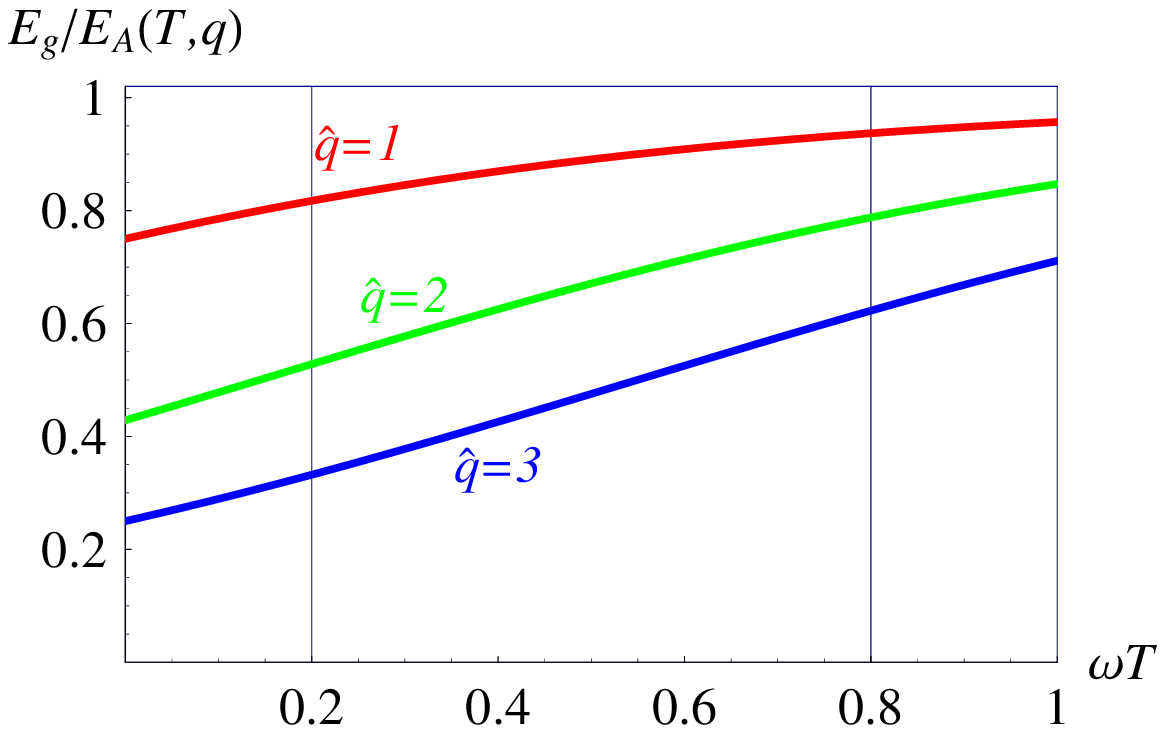}&
\includegraphics[height=.204\textheight,scale=1]{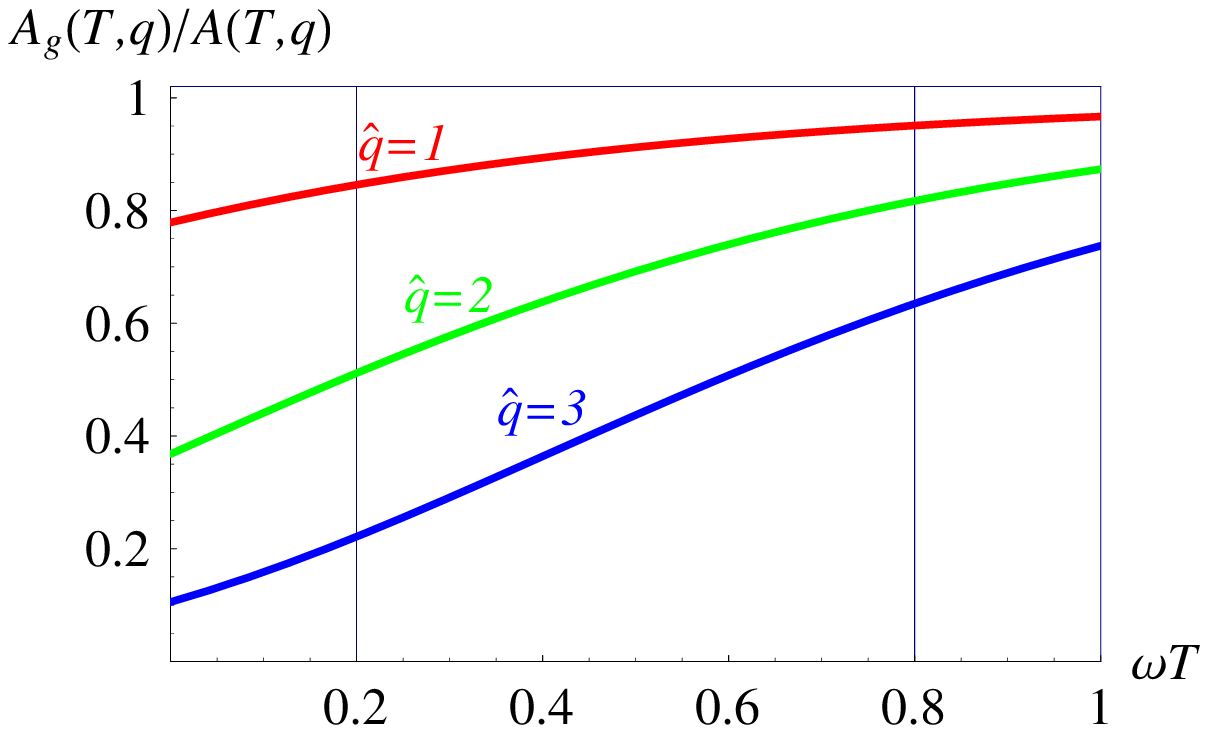}
\end{tabular}\caption{Relative ground-state contributions to
$E_\Gamma(T,q)$ and $\Gamma(T,q)$ (first row), respectively
$E_A(T,q)$ and $A(T,q)$ (second row), for several dimensionless
$\hat q\equiv q/\sqrt{m\,\omega};$ vertical lines delimit the
Borel windows.}\label{Fig:1}\end{figure}

Any such idea is best tested first in a situation in which the
outcome for all bound-state characteristics one is interested in
is known exactly. So let's study a quantum-mechanical model
defined by a nonrelativistic Hamiltonian $H$ with a
harmonic-oscillator interaction:
$$H=\frac{\mathbf{p}^2}{2\,m}+\frac{m\,\omega^2\,r^2}{2}\ ,\qquad
r\equiv|\mathbf{x}|\ .$$This model is {\em exactly\/} solvable:
ground-state energy, decay constant, and form factor
read$$E_\mathrm{g}=\frac{3}{2}\,\omega\ ,\qquad
R_\mathrm{g}\equiv|\psi_\mathrm{g}(\mathbf{x}=\mathbf{0})|^2
=\left(\frac{m\,\omega}{\pi}\right)^{3/2}\ ,\qquad
F_\mathrm{g}(q)=\exp\left(\frac{-q^2}{4\,m\,\omega}\right).$$We
inspect the analogues of a few $N$-point correlation functions of
interpolating currents. The Borelized polarization function or
2-point vacuum--vacuum correlator \cite{LMS:PRD76,LMS:QCD@Work07,
LMS:YaF,LMS:PLB657,LMS:Hadron07,LMS:C8p} reads
$$\Pi(T)\equiv\langle\mathbf{x}_\mathrm{f}=\mathbf{0}|\exp(-H\,T)
|\mathbf{x}_\mathrm{i}=\mathbf{0}\rangle\stackrel{\mathrm{HO}}{=}
\left[\frac{m\,\omega}{2\,\pi\sinh(\omega\,T)}\right]^{3/2}\ .$$
The double-Borelized 3-point correlator of some current operators
$J(\mathbf{q})$ \cite{LMS:PLB671,LMS:C8t} is
given~by\begin{align*}\Gamma(T_2,T_1,q)&\equiv
\langle\mathbf{x}_\mathrm{f}=\mathbf{0}|\exp(-H\,T_2)\,J(\mathbf{q})
\exp(-H\,T_1)|\mathbf{x}_\mathrm{i}=\mathbf{0}\rangle\\[-.09266ex]&
\hspace{-1.845ex}\stackrel{T_{1,2}\to\infty}{\longrightarrow}\quad
R_\mathrm{g}\exp[-E_\mathrm{g}\,(T_1+T_2)]\,F_\mathrm{g}(q)\
;\end{align*}for equal Borel parameters (or Euclidean times)
$T_1=T_2=\frac{T}{2}$ this expression simplifies~to$$\Gamma(T,q)
\stackrel{\mathrm{HO}}{=}\Pi(T)\exp\left[\frac{-q^2}{4\,m\,\omega}
\tanh\left(\frac{\omega\,T}{2}\right)\right]
\quad\stackrel{T\to\infty}{\longrightarrow}\quad
R_\mathrm{g}\exp(-E_\mathrm{g}\,T)\,F_\mathrm{g}(q)
\equiv\Gamma_\mathrm{g}(T,q)\ .$$The Borelized vacuum--hadron
amplitude of the T-product of 2 `quark' currents is
\cite{DM:PLB671,LMS:PRD80}$$A(T,q)\equiv\langle\mathbf{x}=\mathbf{0}|
\exp(-H\,T)\,J(\mathbf{q})|\psi_\mathrm{g}\rangle
\quad\stackrel{T\to\infty}{\longrightarrow}\quad\sqrt{R_\mathrm{g}}
\exp(-E_\mathrm{g}\,T)\,F_\mathrm{g}(q)\equiv A_\mathrm{g}(T,q)\
.$$The admissible working ranges of Borel parameter values, or
`Borel windows', are found by requiring that the ground state
contributes sizeably to the {\em average energies\/} (see
Fig.~\ref{Fig:1})
\begin{align*}E_\Gamma(T,q)&\equiv-\frac{\partial}{\partial
T}\log\Gamma(T,q)=\frac{3}{2}\,\omega\coth(\omega\,T)
+\frac{q^2}{4\,m\left[1+\cosh(\omega\,T)\right]}\ ,\\
E_A(T,q)&\equiv-\frac{\partial}{\partial T}\log
A(T,q)=\frac{3}{2}\,\omega+\frac{q^2}{2\,m}\exp(-2\,\omega\,T)\ ,
\end{align*}and that in the sum rules higher-order nonperturbative
effects are sufficiently suppressed.

\section{Duality $\Longleftrightarrow$ effective continuum
thresholds}
\begin{figure}[ht]\begin{tabular}{cc}
\includegraphics[height=.204\textheight,scale=1]{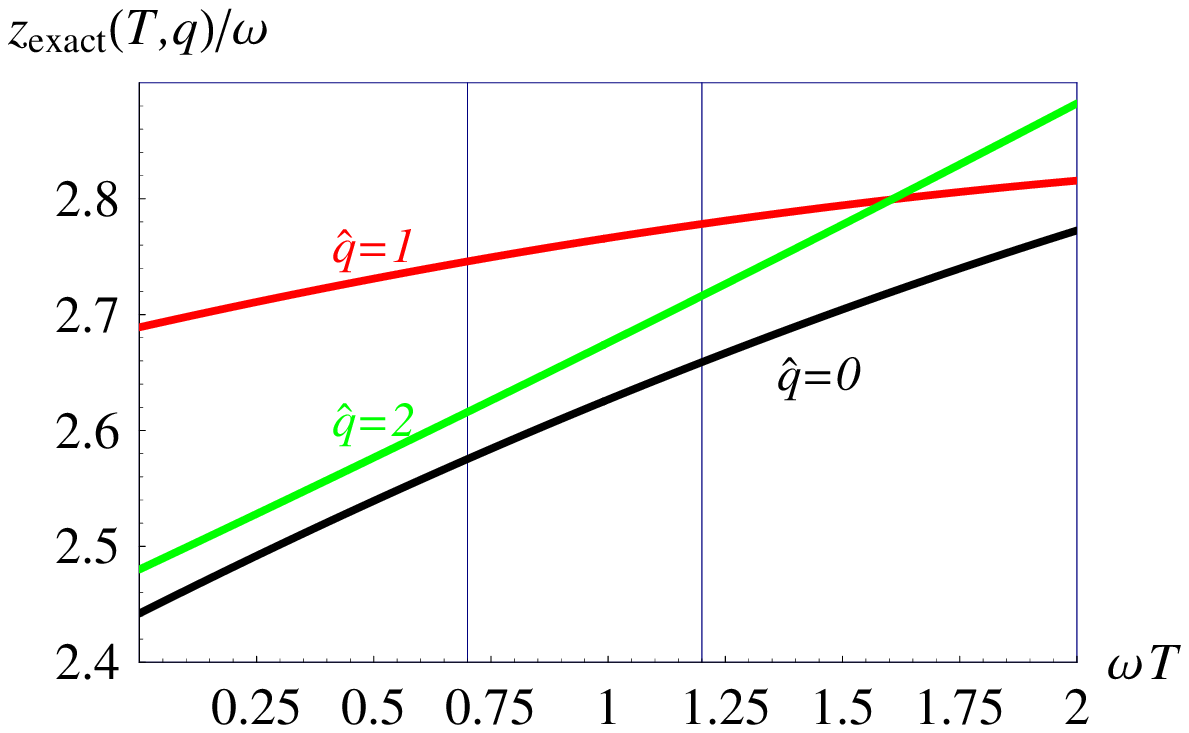}&
\includegraphics[height=.204\textheight,scale=1]{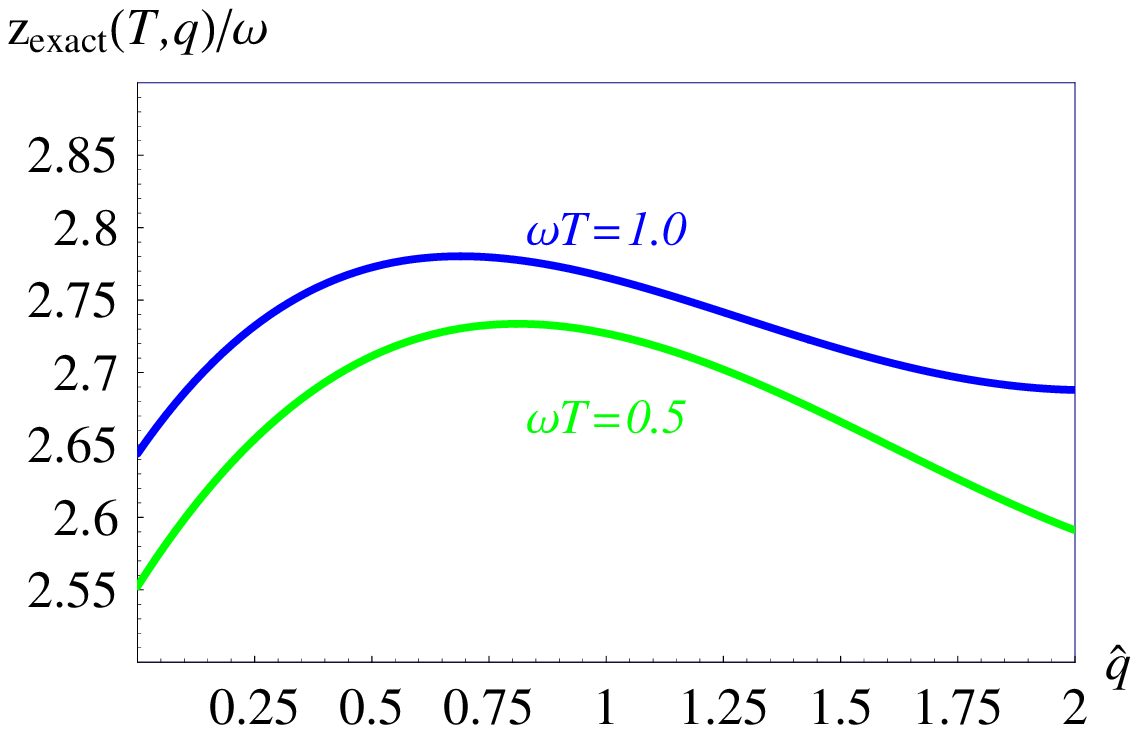}\\
\includegraphics[height=.204\textheight,scale=1]{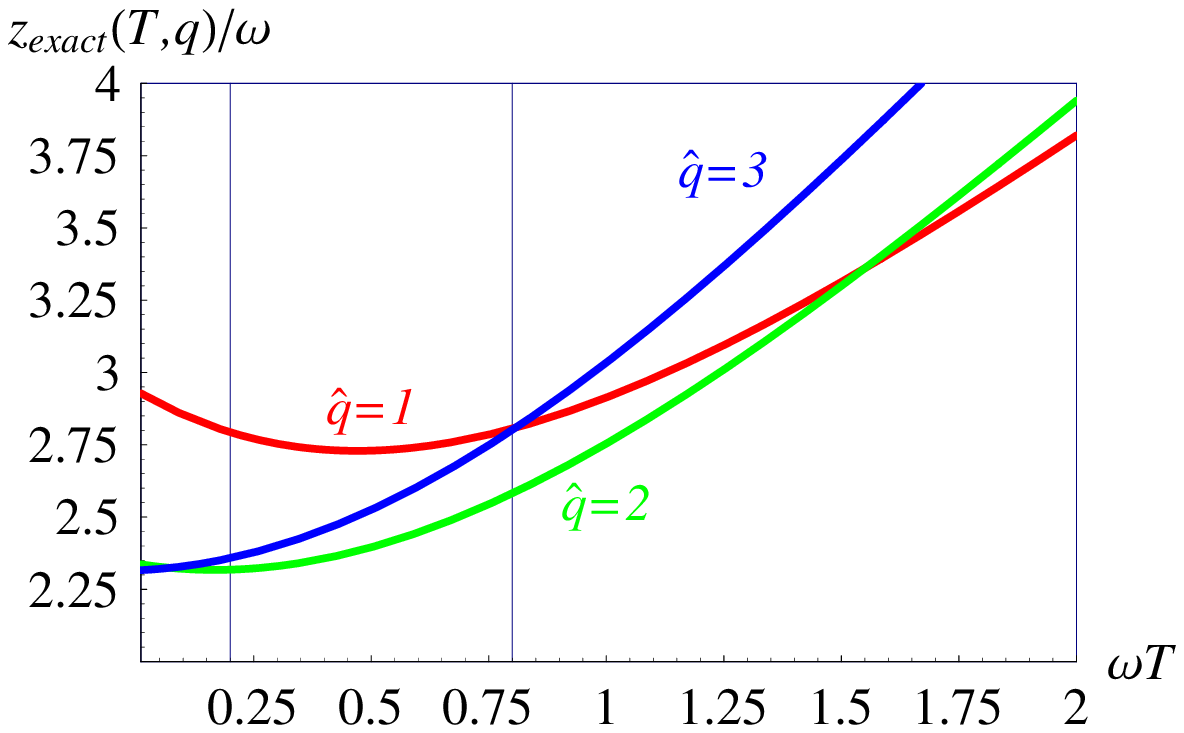}&
\includegraphics[height=.204\textheight,scale=1]{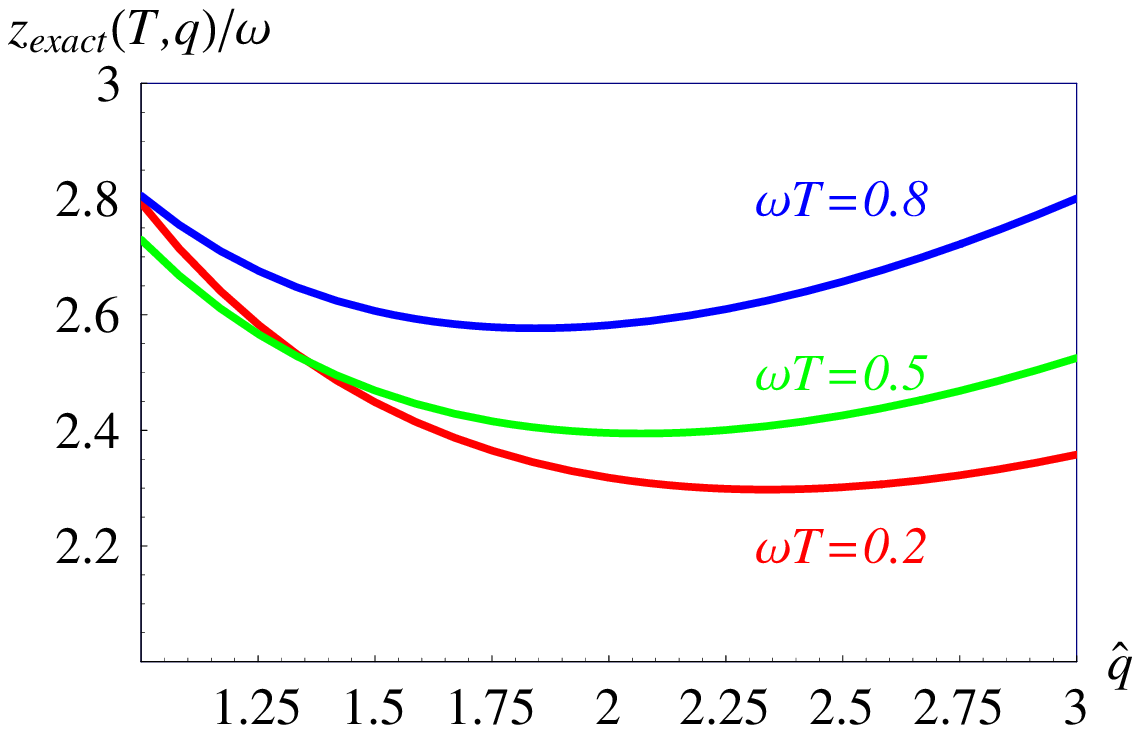}
\end{tabular}\caption{Exact effective continuum thresholds
$z^\Gamma_\mathrm{exact}(T,q)$ (first row) and
$z^A_\mathrm{exact}(T,q)$ (second row) for
$\Gamma_\mathrm{dual}(T,q,z^\Gamma_\mathrm{eff}(T,q))$ and
$A_\mathrm{dual}(T,q,z^A_\mathrm{eff}(T,q))$:
$\Gamma(T_2,T_1,0)=\Pi(T_1+T_2)$ yields
$z^\Gamma_\mathrm{exact}(T,q=0)=z^\Pi_\mathrm{exact}(T).$}
\label{Fig:2}\end{figure}

The derivation of any QCD sum rule proceeds along
several more or less canonical steps:\begin{itemize}\item Perform
the operator product expansion (OPE) of the T-product of involved
fields in terms of local operators yielding so-called vacuum
condensates. In quantum theory, an OPE corresponds to a series
expansion in powers of the Borel parameter
\cite{LMS:PRD76,LMS:QCD@Work07,LMS:YaF, LMS:C8p}.\item Represent
the (perturbative) correlator as dispersion integral over a
spectral density.\item Invoke quark--hadron duality, by cancelling
the contributions to the correlator above its effective continuum
threshold against its hadronic `continuum': we argue that all
thresholds {\em must\/} depend on the Borel parameter and, where
applicable, on momenta.\end{itemize}Even after application of the
quark--hadron duality approximation the resulting sum rules can be
rendered rigorous by introducing the notion of {\em dual
correlators\/}. These QCD-level correlators are defined to be {\em
exactly\/} dual to the relevant lowest, or ground-state, hadronic
contributions. This becomes feasible if allowing the {\em exact
effective continuum thresholds\/} $z_\mathrm{eff}(T,q)$ to depend
on the Borel parameter (and, if relevant, on the
involved~momentum):
\begin{flalign*}&\Pi_\mathrm{dual}(T,z^\Pi_\mathrm{eff}(T))
\equiv\hspace{-1.8ex}\int\limits_0^{z^\Pi_\mathrm{eff}(T)}
\hspace{-1.7ex}\mathrm{d}z\exp(-z\,T)\,\rho_0(z)
+\Pi_\mathrm{power}(T)\stackrel{\mathrm{SR}}{=}
\Pi_\mathrm{g}(T)\equiv R_\mathrm{g}\exp(-E_\mathrm{g}\,T)\
,&\\&\Gamma_\mathrm{dual}(T,q,z^\Gamma_\mathrm{eff}(T,q))
\equiv\hspace{-2.4ex}\int\limits_0^{z^\Gamma_\mathrm{eff}(T,q)}
\hspace{-2.25ex}\mathrm{d}z_1\hspace{-.7ex}
\int\limits_0^{z^\Gamma_\mathrm{eff}(T,q)}
\hspace{-2.25ex}\mathrm{d}z_2\exp\left(-\frac{z_1+z_2}{2}\,T\right)
\Delta_0(z_1,z_2,q)+\Gamma_\mathrm{power}(T,q)&\\[-.7376ex]&
\hspace{17.87ex}\stackrel{\mathrm{SR}}{=}\Gamma_\mathrm{g}(T,q)
\equiv R_\mathrm{g}\exp(-E_\mathrm{g}\,T)\,F_\mathrm{g}(q)\
,&\\&A_\mathrm{dual}(T,q,z^A_\mathrm{eff}(T,q))
\equiv\hspace{-2.4ex}\int\limits_0^{z^A_\mathrm{eff}(T,q)}
\hspace{-2.25ex}\mathrm{d}z\exp(-z\,T)\,\rho_A(z,T,q)
\stackrel{\mathrm{SR}}{=}A_\mathrm{g}(T,q)\equiv\sqrt{R_\mathrm{g}}
\exp(-E_\mathrm{g}\,T)\,F_\mathrm{g}(q)\ .&\end{flalign*}Our exact
knowledge of the quantum-mechanical ground-state mass, decay
constant {\em and\/} form factor renders possible to determine the
exact effective continuum thresholds for all dual correlators:
Fig.~\ref{Fig:2} shows that these thresholds are definitely far
from being constant. However, in order to {\em extract\/} from
$\Gamma_\mathrm{dual}(T,q,z^\Gamma_\mathrm{eff}(T,q))$ or
$A_\mathrm{dual}(T,q,z^A_\mathrm{eff}(T,q))$ the elastic form
factor $F_\mathrm{g}(q)$ by exploiting only the knowledge of
$E_\mathrm{g}$ and $R_\mathrm{g}$ we simultaneously have to pin
down the actual behaviour of the effective continuum thresholds.
We thus adopt for our thresholds the simple polynomial ansatz
\cite{LMS:PRD79,LMS:JPG37,LMS:PRD80}
$z^{(n)}_\mathrm{eff}(T,q)=\sum_{j=0}^nz_j^{(n)}(q)\,(\omega\,T)^j$
and obtain any set of coefficients $z_j^{(n)}(q)$ by evaluating
the {\em dual energies\/}, defined according~to\begin{align*}
E^\Gamma_\mathrm{dual}(T,q)&\equiv-\frac{\mathrm{d}}{\mathrm{d}T}
\log\Gamma_\mathrm{dual}(T,q,z^\Gamma_\mathrm{eff}(T,q))\
,\\E^A_\mathrm{dual}(T,q)&\equiv-\frac{\mathrm{d}}{\mathrm{d}T}\log
A_\mathrm{dual}(T,q,z^A_\mathrm{eff}(T,q))\ ,\end{align*}
\noindent at several Borel-parameter points $T_i,$ $1\le i\le N,$
within the Borel window and minimizing then the squared difference
$\chi^2\equiv\frac{1}{N}\sum_{i=1}^N
\left[E^{\Gamma,A}_\mathrm{dual}(T_i,q)-E_\mathrm{g}\right]^2$
between $E^{\Gamma,A}_\mathrm{dual}(T_i,q)$ and $E_\mathrm{g}.$

\newpage

\begin{figure}[ht]\begin{tabular}{cc}
\includegraphics[height=.204\textheight,scale=1]{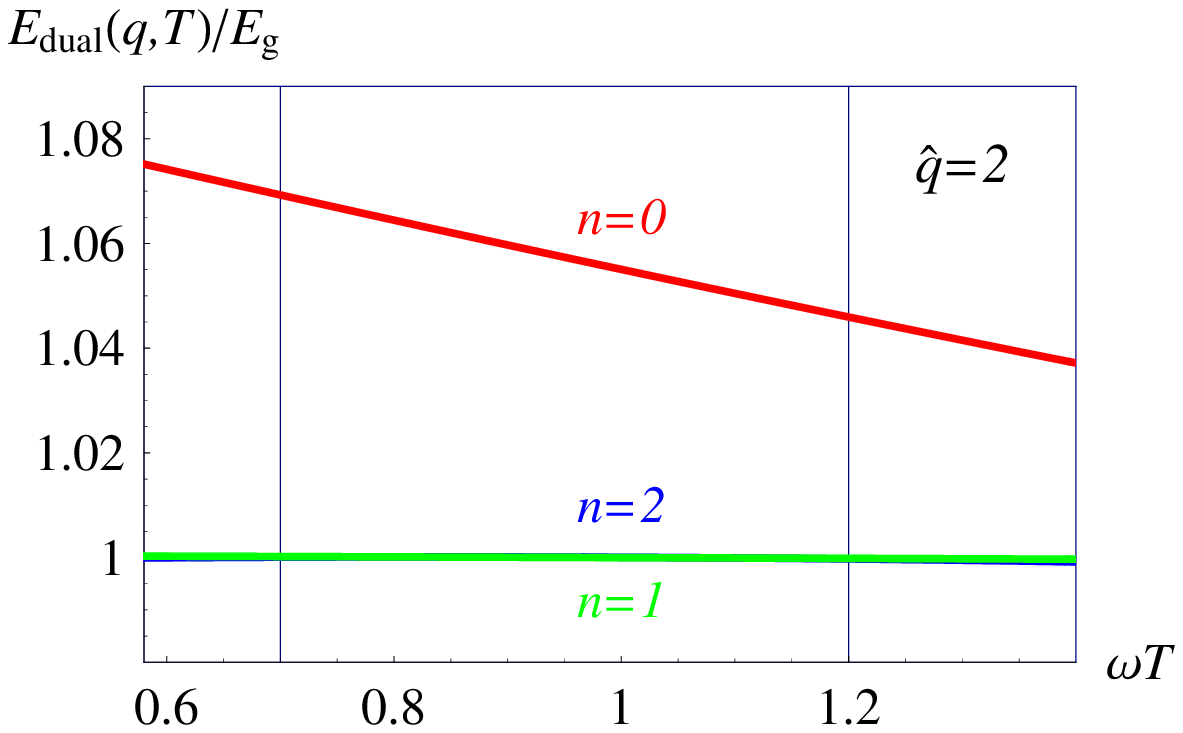}&
\includegraphics[height=.204\textheight,scale=1]{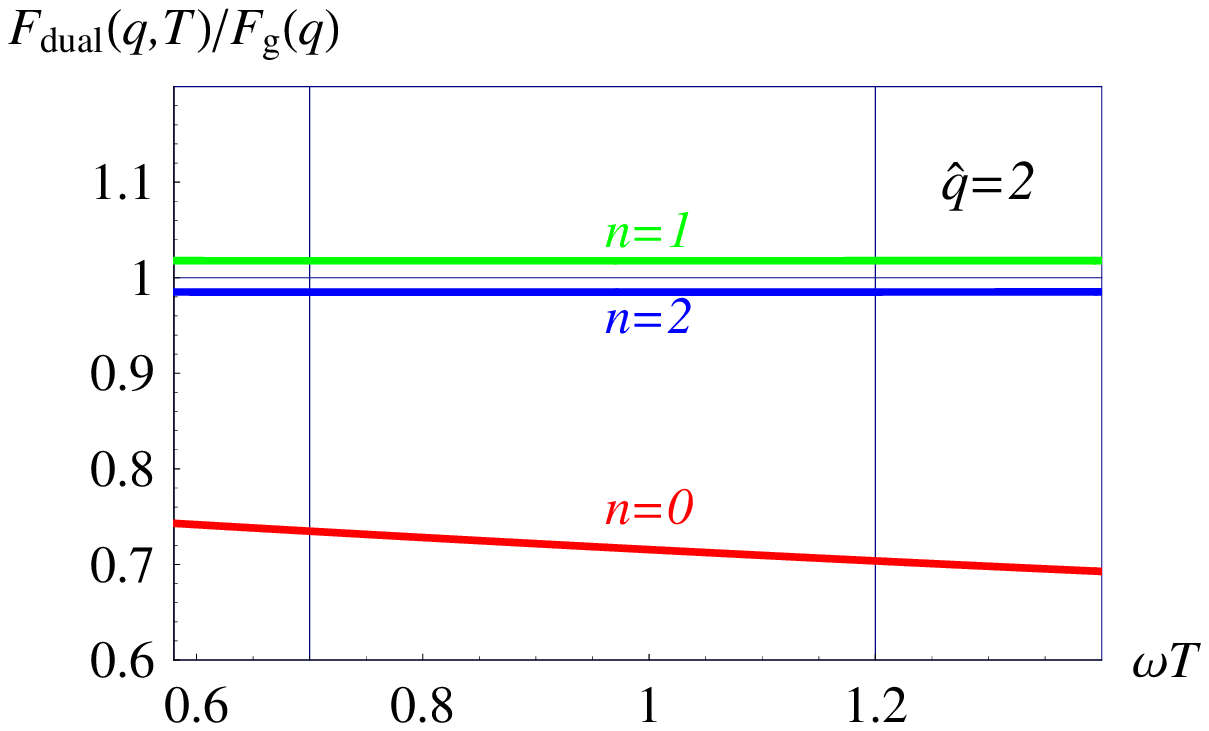}\\
\includegraphics[height=.204\textheight,scale=1]{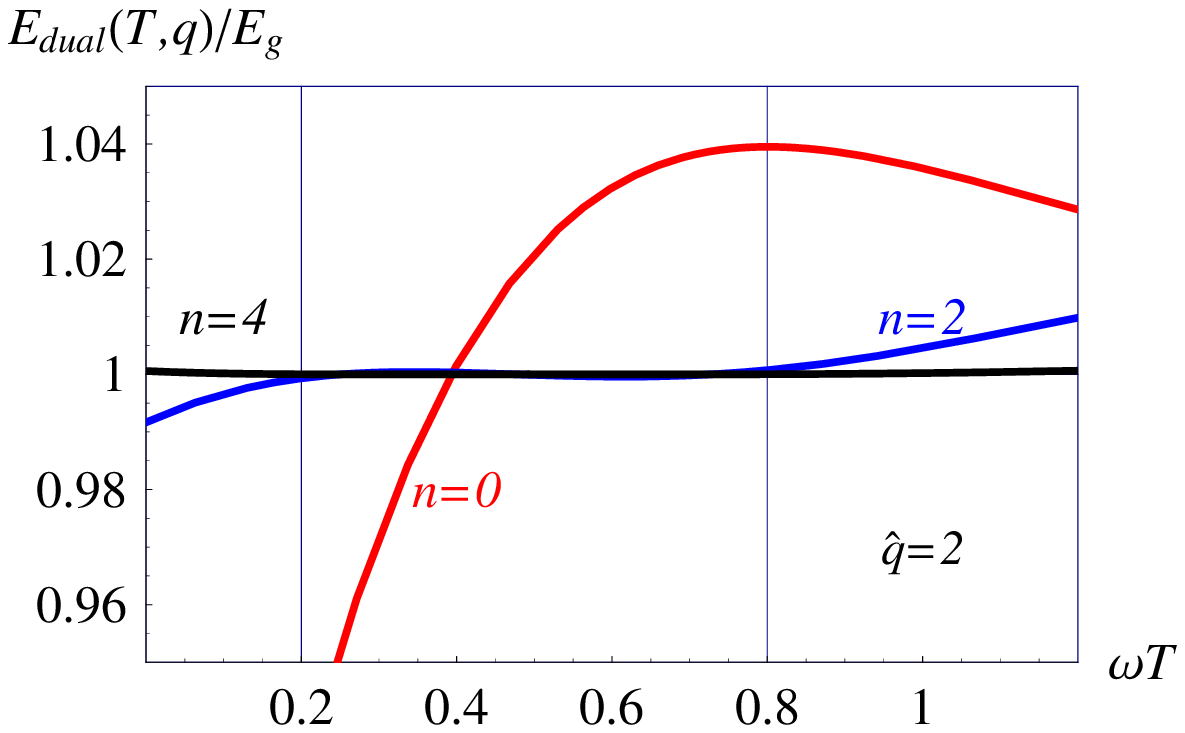}&
\includegraphics[height=.204\textheight,scale=1]{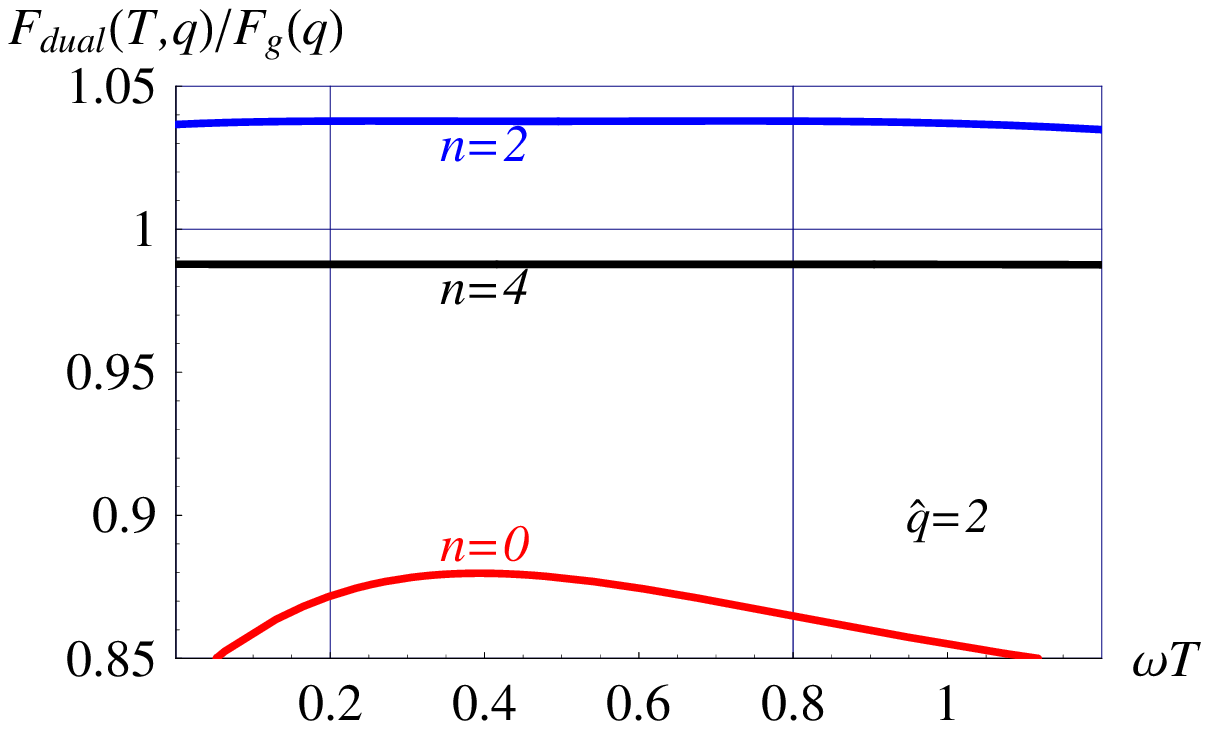}
\end{tabular}\caption{Fitted dual energy $E_\mathrm{dual}(T,q)$
(left) and dual elastic form factor $F_\mathrm{dual}(T,q)$ (right)
for $\Gamma(T,q)$ (upper row) and $A(T,q)$ (lower row) at
dimensionless $\hat q\equiv q/\!\sqrt{m\,\omega}=2.$ These plots
reveal that the intrinsic uncertainties of the sum-rule method
cannot be reliably estimated if assuming a constant threshold
($n=0$).}\label{Fig:4}\end{figure}

\noindent Figure~\ref{Fig:4} depicts the Borel behaviour of our
resulting dual energies and {\em dual form factors\/}.

\section{Bridging the gap: QCD \cite{LMS:PLB687}
$\stackrel{\sim}{\longleftrightarrow}$ potential models
\cite{LMS:YaF2010}}Beyond doubt, our ultimate target must be to
apply our insights on the sum-rule approach gained in the course
of the above quantum-theoretical exercise to QCD in order to
extract characteristics of actual hadrons such as decay constants
and form factors. Let us start by studying the Borelized
correlator of 2 pseudoscalar quark currents
$j_5\equiv(m_\mathrm{b}+m_\mathrm{u})\,\bar
u\,\mathrm{i}\,\gamma_5\,b$:\begin{align*}\Pi(\tau)&=
\hspace{-3.14ex}\int\limits_{(m_\mathrm{b}+m_\mathrm{u})^2}^\infty
\hspace{-3.02ex}\mathrm{d}s\exp(-s\,\tau)\,\rho_\mathrm{pert}(s,\mu)
+\Pi_\mathrm{power}(\tau,\mu)\\&\hspace{-.155ex}
\stackrel{\mathrm{SR}}{=}f_\mathrm{B}^2\,M_\mathrm{B}^4
\exp(-M_\mathrm{B}^2\,\tau)+\cdots
\quad\stackrel{\tau\to\infty}{\longrightarrow}\quad
f_\mathrm{B}^2\,M_\mathrm{B}^4\exp(-M_\mathrm{B}^2\,\tau)
\equiv\Pi_\mathrm{g}(\tau)\ .\end{align*}The perturbative spectral
density $\rho_\mathrm{pert}(s,\mu)$ may be derived from QCD in
form of a~series expansion in powers of the strong coupling
$\alpha_\mathrm{s},$ at an appropriate renormalization
scale~$\mu$:$$\rho_\mathrm{pert}(s,\mu)=\rho_0(s,\mu)+
\frac{\alpha_\mathrm{s}(\mu)}{\pi}\,\rho_1(s,\mu)+
\frac{\alpha_\mathrm{s}^2(\mu)}{\pi^2}\,\rho_2(s,\mu)+\cdots\ .$$

\begin{figure}[h]\begin{tabular}{cc}
\includegraphics[height=.204\textheight,scale=1]{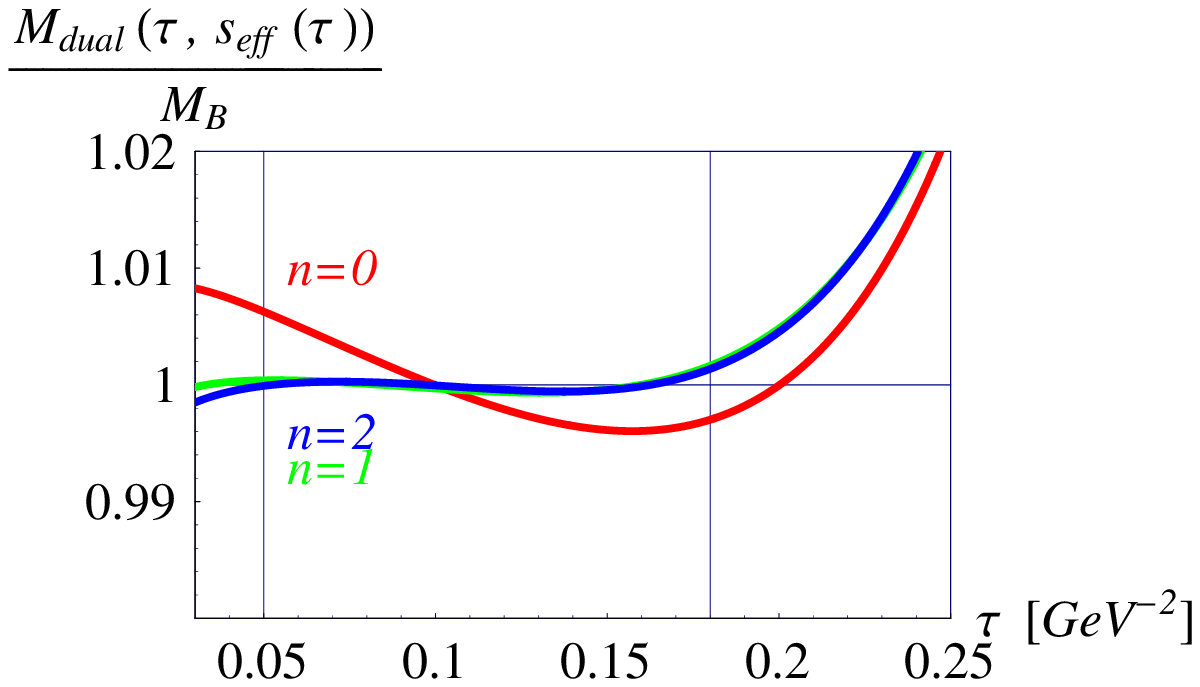}&
\includegraphics[height=.204\textheight,scale=1]{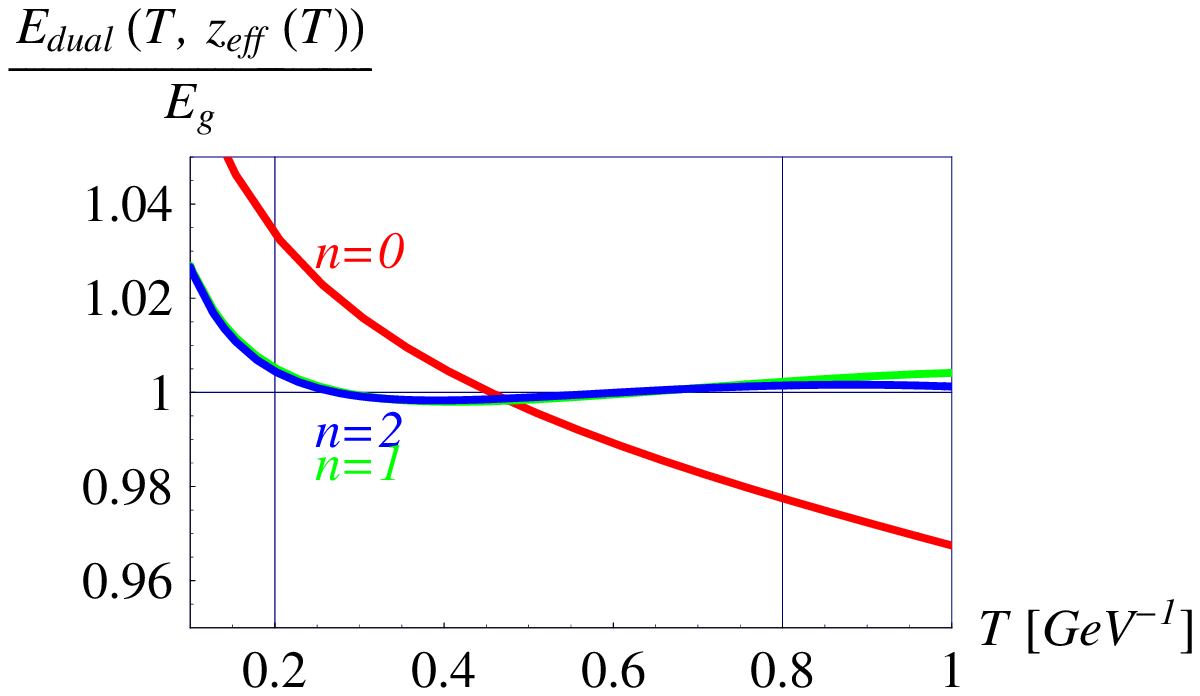}\\
\includegraphics[height=.204\textheight,scale=1]{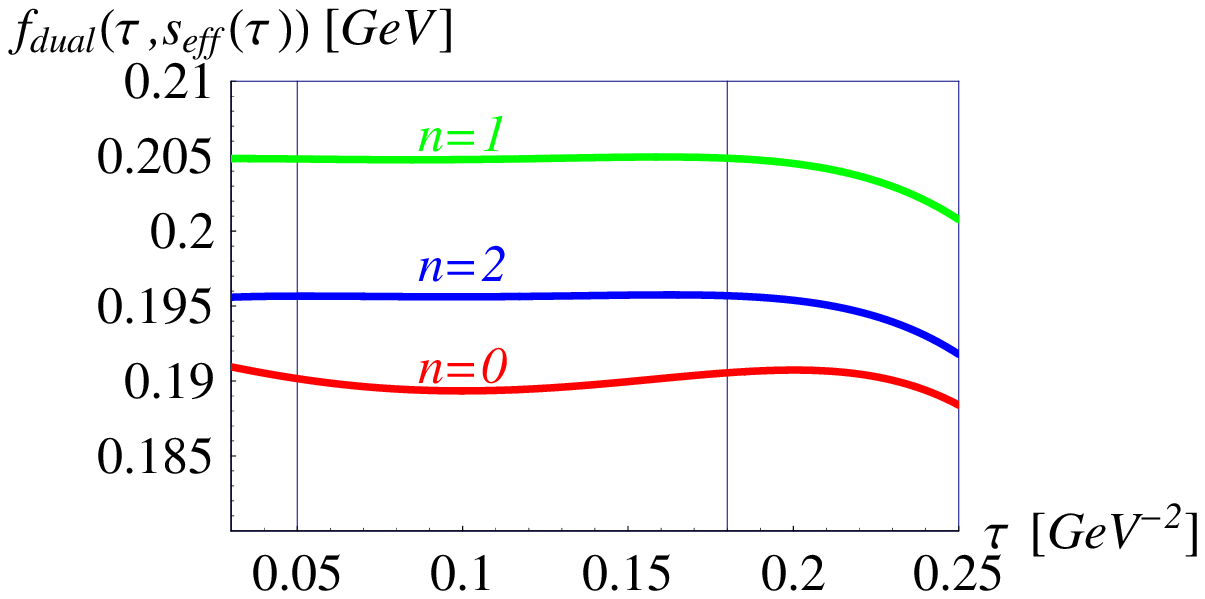}&
\includegraphics[height=.204\textheight,scale=1]{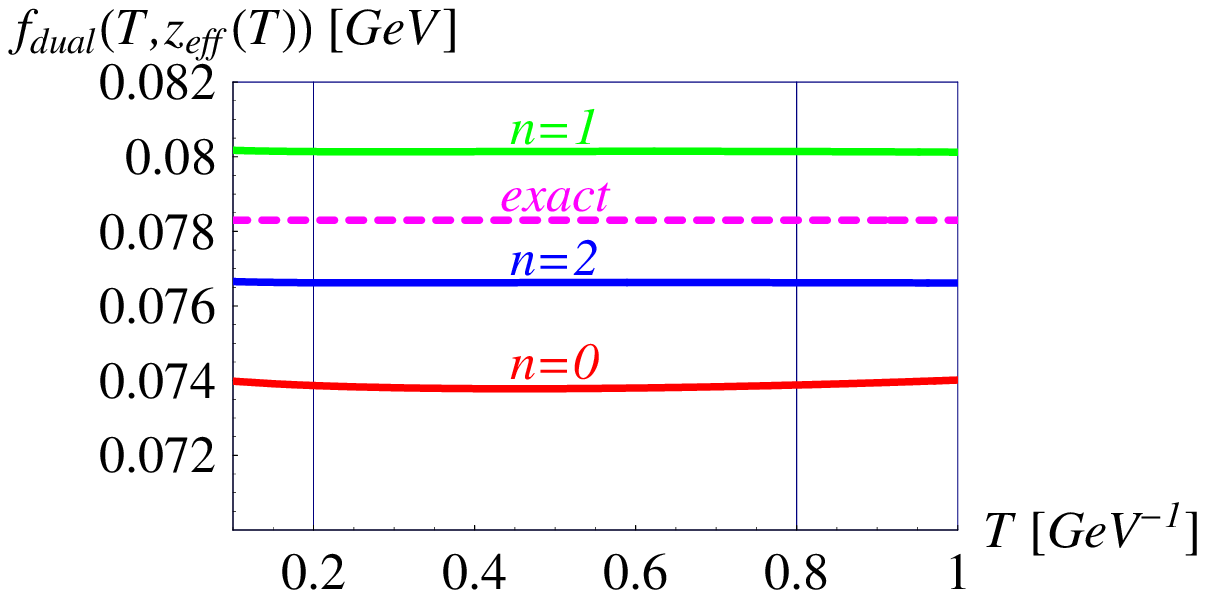}
\end{tabular}\caption{Dual masses and decay constants extracted
from QCD (left) and our potential model
(right).}\label{Fig:5}\end{figure}

\noindent Duality permits dual correlators exactly
counterbalancing all ground-state contributions:\begin{align*}
\Pi_\mathrm{dual}(\tau,s_\mathrm{eff}(\tau))&\equiv\hspace{-3.14ex}
\int\limits_{(m_\mathrm{b}+m_\mathrm{u})^2}^{s_\mathrm{eff}(\tau)}
\hspace{-3.2ex}\mathrm{d}s\exp(-s\,\tau)\,\rho_\mathrm{pert}(s,\mu)
+\Pi_\mathrm{power}(\tau,\mu)\\&\hspace{-.155ex}
\stackrel{\mathrm{SR}}{=}\Pi_\mathrm{g}(\tau)\equiv
f_\mathrm{B}^2\,M_\mathrm{B}^4\exp(-M_\mathrm{B}^2\,\tau)\
.\end{align*}The decay constant $f_\mathrm{B}$ of the charged B
meson is defined by $(m_\mathrm{b}+m_\mathrm{u})\,
\langle0|j_5|\mathrm{B}\rangle=f_\mathrm{B}\,M_\mathrm{B}^2.$ To
parallel the quantum-theoretical case we introduce {\em dual\/} B
mass and decay constant~by
$$M_\mathrm{dual}^2(\tau)\equiv-\frac{\mathrm{d}}{\mathrm{d}\tau}
\log\Pi_\mathrm{dual}(\tau,s_\mathrm{eff}(\tau))\ ,\qquad
f_\mathrm{dual}^2(\tau)\equiv
M_\mathrm{B}^{-4}\exp(M_\mathrm{B}^2\,\tau)\,
\Pi_\mathrm{dual}(\tau,s_\mathrm{eff}(\tau))\ .$$We make a
polynomial ansatz for the {\em effective continuum threshold\/}
$s^{(n)}_\mathrm{eff}(\tau)=\sum_{j=0}^ns_j^{(n)}\,\tau^j$ and get
the coefficients $s_j^{(n)}$ by minimizing
$\chi^2\equiv\frac{1}{N}\sum_{i=1}^N
\left[M^2_\mathrm{dual}(\tau_i)-M_\mathrm{B}^2\right]^2$ for
$\tau_{i\le N}$ in the Borel window \cite{LMS:PLB687}. Figure
\ref{Fig:5} confronts the QCD answer for the B meson with that
from a quantum-theoretical model with harmonic-oscillator plus
Coulomb interaction~potential.

\section{Observations, results, findings, and conclusions}
\begin{enumerate}\item Our exact effective continuum thresholds
{\em do\/} depend on the Borel parameter and any relevant
momentum, and they are {\em not\/} universal but vary with the
studied correlators.\item The algorithm proposed here both
improves significantly the {\em accuracy\/} of traditional
sum-rule predictions and provides reliable estimates of their {\em
intrinsic uncertainties\/}.\item The striking (and surprisingly
even quantitative) {\em similarity\/} of our hadron-parameter
extraction {\em procedures\/} in potential models and in QCD {\em
calls for application in QCD\/}.\end{enumerate}

\begin{theacknowledgments}D.~M.\ has been supported by the Austrian
Science Fund FWF under Project No.~P20573.\end{theacknowledgments}

\bibliographystyle{aipproc}
\begin{thebibliography}{99}
\bibitem{LMS:PRD76}W.~Lucha, D.~Melikhov, and S.~Simula,
\emph{Phys.~Rev.~D\/} \textbf{76}, 036002 (2007) [arXiv:0705.0470
[hep-ph]].
\bibitem{LMS:QCD@Work07}W.~Lucha, D.~Melikhov, and S.~Simula, in
\emph{QCD@Work 2007 --- International Workshop on Quantum
Chromodynamics: Theory and Experiment\/}, edited by P.~Colangelo
et al., AIP Conf.~Proc., AIP, Melville, New York, 2007,
Vol.~\textbf{964}, p.~296 [arXiv:0707.4123 [hep-ph]].
\bibitem{LMS:YaF}W.~Lucha, D.~Melikhov, and S.~Simula,
\emph{Phys.~Atom.~Nucl.\/} \textbf{71}, 1461 (2008).
\bibitem{LMS:PLB657}W.~Lucha, D.~Melikhov, and S.~Simula,
\emph{Phys.~Lett.~B\/} \textbf{657}, 148 (2007) [arXiv:0709.1584
[hep-ph]].
\bibitem{LMS:Hadron07}W.~Lucha, D.~Melikhov, and S.~Simula, in
\emph{XII International Conference on Hadron Spectroscopy ---
Hadron 07\/}, edited by L.~Benussi et al., Frascati Phys.~Ser.,
INFN, Laboratori Nazionali di Frascati, 2007, Vol.~\textbf{46},
p.~1109 [arXiv:0712.0177 [hep-ph]].
\bibitem{LMS:PLB671}W.~Lucha, D.~Melikhov, and S.~Simula,
\emph{Phys.~Lett.~B\/} \textbf{671}, 445 (2009) [arXiv:0810.1920
[hep-ph]].
\bibitem{DM:PLB671}D.~Melikhov, \emph{Phys.~Lett.~B\/}
\textbf{671}, 450 (2009) [arXiv:0810.4497 [hep-ph]].
\bibitem{LMS:C8p}W.~Lucha, D.~Melikhov, and S.~Simula, in
\emph{8\/$^{\it th}$ International Conference on Quark Confinement
and the Hadron Spectrum --- QCHS 2008\/}, PoS (Confinement8), 180
(2009) [arXiv:0811.0533 [hep-ph]].
\bibitem{LMS:C8t}W.~Lucha, D.~Melikhov, and S.~Simula, in
\emph{8\/$^{\it th}$ International Conference on Quark Confinement
and the Hadron Spectrum --- QCHS 2008\/}, PoS (Confinement8), 106
(2009) [arXiv:0811.0721 [hep-ph]].
\bibitem{LMS:PRD79}W.~Lucha, D.~Melikhov, and S.~Simula,
\emph{Phys.~Rev.~D\/} \textbf{79}, 096011 (2009) [arXiv:0902.4202
[hep-ph]].
\bibitem{LMS:JPG37}W.~Lucha, D.~Melikhov, and S.~Simula,
\emph{J.~Phys.~G: Nucl.~Part.~Phys.\/} \textbf{37}, 035003 (2010)
[arXiv:0905.\ 0963 [hep-ph]].
\bibitem{LMS:PRD80}W.~Lucha, D.~Melikhov, H.~Sazdjian, and
S.~Simula, \emph{Phys.~Rev.~D\/} \textbf{80}, 114028 (2009)
[arXiv:0910.\ 3164 [hep-ph]].
\bibitem{LMS:PLB687}W.~Lucha, D.~Melikhov, and S.~Simula,
\emph{Phys.~Lett.~B\/} \textbf{687}, 48 (2010) [arXiv:0912.5017
[hep-ph]].
\bibitem{LMS:YaF2010}W.~Lucha, D.~Melikhov, and S.~Simula,
arXiv:1003.1463 [hep-ph], \emph{Phys.~Atom.~Nucl.\/} \textbf{73}
(in print).\end{thebibliography}
\end{document}